\documentclass[twocolumn,showpacs,amsmath,amssymb]{revtex4}
\usepackage[dvips]{graphicx}
\usepackage{bm}

\begin{document}
\title{Indirect Exchange Interaction
between two Quantum Dots in an
Aharonov-Bohm Ring}
\author{Yasuhiro Utsumi$^1$,
Jan Martinek$^{2,3}$,
Patrick Bruno$^1$,
and
Hiroshi Imamura$^4$}
\address{
$^1$
Max-Planck-Institut f\"ur Mikrostrukturphysik
Weinberg 2, D-06120 Halle (Saale), Germany\\
$^2$
 Institut f\"{u}r
Theoretische Festk\"{o}perphysik, Universit\"{a}t Karlsruhe, 76128
Karlsruhe, Germany \\ $^3$ Institute of Molecular Physics, Polish
Academy of Sciences, 60-179 Pozna\'n, Poland\\
 $^4$ Graduate School of Information
Sciences, Tohoku University, Sendai 980-8579, Japan}
%

%
%
\begin{abstract}
We investigate the Ruderman-Kittel-Kasuya-Yosida (RKKY)
interaction between two spins located at two quantum dots
embedded in an Aharonov-Bohm (AB) ring. In such a system the RKKY
interaction, which oscillates as a function of the distance
between two local spins, is affected by the flux.
For the case of the ferromagnetic RKKY interaction,
we find that the amplitude of AB oscillations is enhanced by
the Kondo correlations
and an additional maximum appears at half flux,
where the interaction is switched off.
For the case of the antiferromagnetic RKKY interaction, we find
that the phase of AB oscillations is shifted by $\pi$, which is
attributed to the formation of a singlet state between two
spins for the flux value close to integer value of flux.
\end{abstract}

\date{\today}
\maketitle

\newcommand{\rd}{d}
\newcommand{\ri}{i}
\newcommand{\mat}[1]{\mbox{\boldmath$#1$}}

\section{introduction}

When two magnetic moments are embedded in a metal, they induce
spin polarization in a conduction electron sea and couple each
other even they are spatially apart. Such indirect exchange
interaction, the Ruderman-Kittel-Kasuya-Yosida (RKKY) interaction,
has been known from the 1950s \cite{Kittel}. The indirect exchange
interaction in magnetic nanostructures is one of basic mechanisms
for spintronics \cite{maekawaBook} and it is well understood for
ferromagnet and nonmagnetic metal multilayer structures
\cite{Bruno}. However for semiconductor nanostructures, the
indirect exchange interaction between local spins formed in two
quantum dots has not yet been observed, in spite of the importance
as a basic physics and the potential application for semiconductor
nanospintronics.

Recent improvement of the fabrication technique for semiconductor
nanostructures enables one to make rather complicated structures
with possibility of the precise controlling of their parameters.
For example, a double quantum dot (QD) system \cite{Holleitner}
and the composite system of QD and an Aharonov-Bohm (AB) ring have
been made \cite{Yacoby,Kobayashi1}.
The double-dot system was proposed for a candidate of the \lq \lq
qubit", because in the Coulomb blockade (CB) regime a dot with odd
numbers of electrons, behaves as a local spin and two dot spins
can be entangled by introducing the exchange interaction between
them \cite{Loss}.
Such exchange interaction has been also discussed from the point
of view of the competition between the Kondo effect and the
antiferromagnetic (AF) interaction \cite{Georges}. However the
direct exchange interaction was considered rather than the RKKY
interaction.
Investigations on the AB ring embedded with QD are aimed at
understanding the coherent transport through QD
\cite{Yacoby,Konig_1,Konig_2} and the indirect exchange
interaction between two local spins has not been addressed.

So it is intriguing to investigate the RKKY interaction between
two QDs in CB regime embedded in AB ring. We will show that the
RKKY interaction, the sign of which oscillate as a function of the
distance (RKKY oscillations), is affected by the flux and it
dominates the transport properties. For ferromagnetic (F) coupling
between dot spins, the amplitude of AB oscillations is enhanced by
Kondo correlations and an additional maximum appears at half flux.
For AF coupling case,
the phase of AB oscillations is shifted by $\pi$.

\section{model and calculations}


\begin{figure}[b]
\includegraphics[width=0.60\columnwidth]{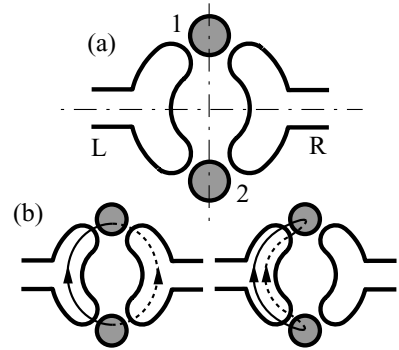}
\caption{ (a) Aharonov-Bohm ring embedded with one QD in each arm.
The system has the parity symmetry along the horizontal and
vertical axes (dot-dashed lines). (b) The flux dependent (left
panel) and independent (right panel) particle-hole excitation.
$\Phi$ is the flux penetrating the ring. The directed solid and
dashed lines are the particle and hole propagators, respectively.
} \label{fig:ABring}
\end{figure}
Figure \ref{fig:ABring}(a) shows the schematic picture of
an AB ring embedded with one QD in each arm.
QDs denoted with 1 and 2 weakly couple to left and right leads.
The effective Hamiltonian is written with
two single-channel 1-dimensional (1D) leads Hamiltonian $H_0$
and
the tunneling Hamiltonian $H_{\rm T}$
as $H=H_0+H_{\rm T}$.
The lead Hamiltonian is given by
$
H_0
=
\sum_{k \, r=L,R \, \sigma}
\varepsilon_k
a_{k r \sigma}^{\dag} a_{k r \sigma}
$,
where $a_{k L(R) \sigma}$ is an annihilation operator of an electron
with quantum number $k$ and spin $\sigma$ in the
left (right) lead.
For simplicity, we adopt the so-called
Coqblin-Schrieffer model,
\begin{equation}
H_{\rm T}
=
\sum_{\stackrel{\scriptstyle{r,r'=L,R}}{n=1,2}}
\sum_{\sigma,\sigma'=\uparrow,\downarrow}
\frac{J}{2}
a_{r n \sigma}^{\dagger}
X^n_{\sigma' \, \sigma}
a_{r' n \sigma'},
\label{eqn:HT}
\end{equation}
for the tunneling Hamiltonian through QDs with odd numbers of
electrons in CB regime. The Hubbard operator $X^n_{\sigma' \,
\sigma}=| n, \sigma' \rangle \langle n, \sigma|$ describes the
spin state of the $n$-th QD and $J>0$ is a coupling constant. The
annihilation operator $a_{r n \sigma}$ is written using the
projection $\langle n| k r \rangle$ of wave function of an
electron in the lead $r$ with quantum number $k$ at the boundary
of the $n$-th QD, as
$
a_{r n \sigma}
\!\!
=
\!\!
\sum_k
\langle n| k r\rangle
\,
a_{k r \sigma}
$
\cite{Bruder}.

Here we encounter a problem: One needs to know the proper wave
function including the information on the coherent propagation of
an electron through arms. Usually the scattering theory
\cite{Gefen} is suitable for treating the electron coherency.
However it is complicated to combine this theory with a theory
based on the Hamiltonian in the second-quantization
representation.
In this paper we circumvent this problem.
Rather, we utilize an assumption of
parity (mirror) symmetry
along the horizontal and the vertical axes
(dot-dashed lines in Fig. \ref{fig:ABring}(a)).
Namely, the total Hamiltonian is invariant
under the interchange of indices
$L \leftrightarrow R$
or
$1 \leftrightarrow 2$.
Such approximation will make calculations simple and will contain
all important physics. Any deviations from such a symmetry will
change the result only quantitatively.

For practical calculations, it is convenient to introduce an
annihilation operator of even/odd parity states \cite{Jayaprakash}
$
a_{r \pm \sigma}
\!
=
\!
(a_{r 1 \sigma} \! \pm \! a_{r 2 \sigma})
/\sqrt{2}
$,
(
$
a_{r \pm \sigma}
\!\!
=
\!\!
\sum_k
\langle \pm | k \rangle
\,
a_{k r \sigma}
$,
where $| \pm \rangle \!=\! (| 1 \rangle \! \pm \! | 2 \rangle)/\sqrt{2}$ ;
we dropped the index $r$ from $| k r\rangle$
because of the parity symmetry along the vertical axis).
Annihilation operators for even and odd parity states are orthogonal,
\begin{equation}
\left \{
a_{r p \sigma},
a_{r' p' \sigma'}
\right \}
=
\delta_{rr'}
\delta_{\sigma \sigma'}
\delta_{pp'},
\nonumber
\end{equation}
because of the parity symmetry along the horizontal axis.

When magnetic field is applied, an AB phase factor ${\rm e}^{\ri
\phi/2}$ (${\rm e}^{-\ri \phi/2}$) must be counted in
Eq.~(\ref{eqn:HT}), when electron tunnels through a QD in the
clockwise(anticlockwise) direction \cite{Hackenbroich}. The AB
phase is written with vector potential $\vec{A}$ as,
\begin{equation}
\phi=
\frac{2 \pi}{\Phi_0}
\oint \vec{A} \cdot d \vec{l},
\label{eqn:ABphase}
\end{equation}
where $\Phi_0= h c/e$ is the flux quantum
and the line integral is performed along the
ring in the clockwise direction.
An AB flux breaks the time-reversal symmetry
and it generates the main difference between
features of the orthodox two-impurity Kondo model
\cite{BealMonod,Jayaprakash,Jones1,Jones2,Jones3}
and the AB ring embedded with one QD in each arm.
Here, we note that the
magnetic field in leads and QDs is not zero
for an experiment and it causes
Zeeman splitting of electron spins.
In the following discussions, we consider an
ideal situation where there is no Zeeman splitting
and discuss the effect of AB phase on
the indirect exchange coupling and transport properties.


The Hamiltonian for the RKKY interaction can be obtained by the
second order perturbation theory in terms of $J/\varepsilon_{\rm
F}$, where $\varepsilon_{\rm F}$ is the Fermi energy
\cite{Kittel,Schwabe}:
\begin{equation}
H_{\rm RKKY}= \frac{J_{\rm RKKY}(\phi)}{2} \sum_{\sigma \sigma'}
X^1_{\sigma \sigma'} X^2_{\sigma' \sigma} \; .
 \label{eqn:Ha_RKKY}
\end{equation}
The coupling constant $J_{\rm RKKY}(\phi)$ can be written as
\begin{equation}
J_{\rm RKKY}(\phi)
=
\frac{J^2}{2}
\, \chi \,
(2+2 \cos \phi),
\label{eqn:defRKKY}
\end{equation}
where a succeptibility function $\chi$ can be found by the
perturbation theory based on the Keldysh Green function technique
\cite{Schwabe}.
In the equilibrium, it can be written as
\begin{eqnarray}
J^2
\chi
\! \!
&=&
\! \!
\frac{1}{4}
\,
{\rm Re}
\! \!
\int
\rd \varepsilon
\rd \varepsilon'
\,
\frac{
\gamma_p(\varepsilon)
\gamma_p(\varepsilon')
-
\gamma_p(\varepsilon) \gamma_{\bar{p}}(\varepsilon')
}{\varepsilon+ \ri \eta-\varepsilon'} \nonumber \\ &\times& \!\!\!
\left \{ f^{+}(\varepsilon)-f^{+}(\varepsilon') \right \} \; ,
\label{eqn:chi}
\end{eqnarray}
where $\eta$ is a positive infinitesimal number
and
$
\gamma_p(\varepsilon)
=
J
\sum_k
\langle p| k \rangle
\langle k | p \rangle
\,
\delta (\varepsilon-\varepsilon_k)
$
is a spectral function of parity $p$ \lq \lq electron propagator".
The subscript $\bar{p}$ represents the opposite parity of $p$,
i.e. $\bar{p}=\pm$ for $p=\mp$. Here,
$f^{\pm}(\varepsilon)=1/(1+{\rm e}^{\pm \beta \varepsilon})$ is
the electron (hole) Fermi distribution function, and $\beta \equiv
T^{-1}$(We use the unit $k_{\rm B} \equiv 1$).
In Eq. (\ref{eqn:defRKKY}), a phase dependent factor $( 2 + 2 \cos
\phi)$ appears, which value is related to the four configurations
of particle-hole excitations - two of which enclose the flux [left
panel of Fig. \ref{fig:ABring}(b)] and pick up a phase factor
${\rm e}^{\ri \phi}$ or ${\rm e}^{-\ri \phi}$ and give term $2
\cos \phi$, and the others (right panel) are independent of the
flux - give term $2$. The Eq. (\ref{eqn:defRKKY}) is one of the
main results of this paper. It shows that by means of external
flux $ \phi$ one can control the amplitude of the RKKY interaction
but it is impossible to change its sign since $( 2 + 2 \cos \phi)
\geq 0$.

Since we consider 1D leads, we approximate $\gamma_p(\varepsilon)$
as for the 1D free electron gas with the linearized dispersion
relation \cite{Jayaprakash}:
\begin{equation}
\gamma_{\pm}(\varepsilon)
 \simeq
 \bar{J} \left [ 1 \pm \cos \left
\{ k_{\rm F} l \left ( 1 +\frac{\varepsilon}{D} \right ) \right \}
\right ], \label{eqn:green}
\end{equation}
where $k_{\rm F}$ is the Fermi wave number and $l$ is the length
of an electron path between two QDs. The argument of cosine
function is the energy dependent \lq \lq orbital phase"
\cite{kubala_2}, i.e. the accumulated phase during electron
propagation between two QDs. We introduced the cut-off energy
$D=\hbar v_{\rm F} k_{\rm F}$, where $v_{\rm F}$ is the Fermi
velocity. Here, $\bar{J}$ is written with the density of the
states $\rho$ as $\bar{J} \equiv J \rho$. Substituting Eq.
(\ref{eqn:green}) into Eqs. (\ref{eqn:defRKKY}) and
(\ref{eqn:chi}), we obtain for short distance between two QDs
($k_{\rm F} l \ll 2 \pi$) the ferromagnetic coupling as
$
J_{\rm RKKY}(\phi)
 \simeq
-2 \ln 2 \, \bar{J}^2 D
\,
( 2+2 \cos \phi )
$
and more relevant - the RKKY oscillations of 1D free electron gas
\cite{Kittel,Yafet} as a function of $l$ for long distance between
dots ($k_{\rm F} l \gg 2 \pi$)
\begin{equation}
J_{\rm RKKY}(\phi)
\simeq
 -\frac{\pi \bar{J}^2 D \cos(2 k_{\rm F}
l)}{4 \, k_{\rm F} l} \, (2+2\cos \phi). \label{eqn:RKKYapprox}
\end{equation}
Above expressions are obtained by replacing the Fermi functions in
Eq. (\ref{eqn:chi}) with those at $T \! = \! 0$, what is
valid below the characteristic temperature $T^*$ defined by,
\begin{equation}
T^* \equiv \frac{\hbar}{\tau},
\; \;
\tau \equiv \frac{\hbar k_{\rm F} l}{D}=\frac{l}{v_{\rm F}}.
\label{eqn:characteristictime}
\end{equation}
Here, $\tau$ is the characteristic time scale for
an electron travels between two QDs.
It can be understood from the following argument: Electrons deep
inside the Fermi sea are responsible for the RKKY oscillations. On
the other hand, electrons with energy $\varepsilon$
($|\varepsilon| \ll T^*$), i.e. electrons around the Fermi level,
are unimportant, because such electrons contribute only
oscillations whose characteristic wave length $h v_{\rm F}/T^*$ is
much longer than $l$. Thus the RKKY oscillations are insensitive
to the temperature in the regime $T \ll T^*$. However, when the
temperate reaches $T^*$, the RKKY oscillations are affected by the
thermal excitations of lead electrons and will be smeared out.

Due to the RKKY interaction, depending on the sign of the coupling
$J_{\rm RKKY}(\phi)$ [Eq. (\ref{eqn:RKKYapprox})], the two dot
spins are entangled and form a singlet state $|0,0 \rangle$ for AF
coupling ($J_{\rm RKKY}(\phi) > 0$) or a triplet state $|1,m
\rangle$ ($m=0,\pm 1$) for F coupling ($J_{\rm RKKY}(\phi) < 0$).

In the following, we will discuss how the flux dependent RKKY
interaction modifies transport properties of the ring. To
calculate the conductance, we adopt the third order perturbation
theory in terms of $\bar{J}$, in order to take into account the
Kondo correlations: They are pronounced when the temperature
decreases and approaches the order of the Kondo temperature
$T_{\rm K} \! \sim \! D \exp(-1/(2 \bar{J}))$. In our case, the
RKKY interaction becomes important at the temperature, where the
Kondo correlations are also unignorable.
First we rewrite the tunnel Hamiltonian, Eq.~(\ref{eqn:HT}), using
the vector operator of the $n$-th local spin $\vec{S}^n$, whose
components are defined as $S^n_{+/-}=X^n_{\uparrow \downarrow /
\downarrow \uparrow}$ and $S^n_z=(X^n_{\uparrow
\uparrow}-X^n_{\downarrow \downarrow})/2$.
Further we introduce operators
$\vec{S}^{\pm}=\vec{S}^1 \pm \vec{S}^2$,
which satisfy the following commutation relations:
\begin{equation}
[S^p_i,S^p_j]=\ri \epsilon_{ijk} S^+_k,
\; \;
[S^p_i,S^{\bar{p}}_j] = \ri \epsilon_{ijk} S^-_k,
\label{eqn:commute}
\end{equation}
where $\epsilon_{ijk}$ is the Levi-Chivita antisymmetric tensor.
The operator $\vec{S}^{+}$ does not change the total spin-quantum
number, while $\vec{S}^{-}$ is the operator of the singlet-triplet
transition. By using above operators, we obtain the symmetrized
form of $H_{\rm T}$ \cite{BealMonod,Matho,Jayaprakash} taking into
account the AB phase as
\begin{eqnarray}
H_{\rm T} \!\!
&=& \!\!
\frac{J}{4} \!\!\!
\sum_{\stackrel{\scriptstyle{p=\pm}}{\sigma,\sigma'=\uparrow,\downarrow}}
\left \{
\sum_{r}
\left(
\vec{\sigma}_{r r}^p \!\!
\cdot
\vec{S}^p
+
v
\,
a_{r p \sigma'}^{\dagger}
\delta_{\sigma' \sigma}
a_{r p \sigma}
\right)
\right.
\nonumber \\
&+& \!\!\!
\cos \frac{\phi}{2}
\left(
\vec{\sigma}_{RL}^p \!\!
\cdot
\vec{S}^p
+
v \,
a_{R p \sigma'}^{\dagger}
\delta_{\sigma' \sigma}
a_{L p \sigma}
\right)
\!\!+\!
{\rm H. c.}
\nonumber \\
&+& \!\!\!
\left.
\ri \sin \frac{\phi}{2}
\left(
\vec{\sigma}_{RL}^{\bar{p}} \!\!
\cdot
\vec{S}^p
+
v \, a_{R \bar{p} \sigma'}^{\dagger}
\delta_{\sigma' \sigma} a_{L p \sigma}
\right)
\!\!+\!
{\rm H. c.}
\right \}.
\nonumber \\
\label{eqn:symmetrizedHT}
\end{eqnarray}
Here,
$\vec{\sigma}^{+(-)}_{r' r}
=
\sum_{\sigma \sigma' p} a_{r' p \sigma'}^{\dagger}
\vec{\sigma}_{\sigma' \sigma} a_{r p(\bar{p}) \sigma}$
denotes effective conducting electron spin and
 is
defined with the vector Pauli matrix $\vec{\sigma}$. Terms
proportional to $v$ represent potential scattering process and for
our case, $v=1$.
The first line represents the reflection process
and shows that the change of parity and
the singlet-triplet transition occur simultaneously.
The second and the third lines describe transmission processes.
The third line describes the singlet-triplet transition without
changing the parity, that is not invariant under the interchange
of indices, $L \leftrightarrow R$ or $1 \leftrightarrow 2$, (i.e.
the replacement of $a_{{\rm r} \pm \sigma}$ with $\pm a_{{\rm r}
\pm \sigma}$).
Here it does not mean that the parity symmetry is broken: the space
inversion transformation changes $\phi \rightarrow -\phi$, because
it also reverses the direction of the line integral in Eq.
(\ref{eqn:ABphase}).

In order to calculate the linear conductance, we adopt the
diagrammatic technique for the density matrix in the real-time
domain \cite{Schoeller_Schon,Konig1}. With the help of the
commutation relations, Eq. (\ref{eqn:commute}), the perturbative
calculation is performed rather systematically
(Appendix.~\ref{appendix:perturbation}). The \lq \lq partial
self-energy" which represents the transition rate for an electron
from the left lead to the right lead accompanied by the
triplet-triplet transition $\Sigma^{LR}_{11}$ or preserving a
singlet state $\Sigma^{LR}_{00}$ or accompanied by the
singlet-triplet transition $\Sigma^{LR}_{j \bar{j}}$ ($j=0,1$) is
obtained as follows:
\begin{widetext}
\begin{eqnarray}
\Sigma^{LR}_{11}
&\simeq&
\frac{3 \pi \ri}{2}
\,
\int \rd \varepsilon
\sum_{p=\pm}
\left[
\gamma^+_{p L}(\varepsilon)
\,
\gamma^-_{p R}(\varepsilon)
\,
{\rm Re}
\left \{
1+\frac{v^2}{2}
+
\sigma_{1 \, p}(\varepsilon)
+
\sum_{j=0,1}
\frac{
\sigma_{1 \, \bar{p}}(\varepsilon-\Delta_{j \bar{j}})
}{2}
\right \}
\cos^2 \frac{\phi}{2}
\right.
\nonumber \\
&+&
\left.
\gamma^+_{p L}(\varepsilon)
\,
\gamma^-_{\bar{p} R}(\varepsilon)
\,
{\rm Re}
\left \{
1+\frac{v^2}{2}
+
\sigma_{0 \, p}(\varepsilon)
+
\sum_{j=0,1}
\frac{
\sigma_{0 \, \bar{p}}(\varepsilon-\Delta_{j \bar{j}})
}{2}
\right \}
\sin^2 \frac{\phi}{2}
\right],
\nonumber \\
\label{eqn:tt}
\\
\Sigma^{LR}_{00}
&\simeq&
\frac{\pi \ri}{4}
\,
\int \rd \varepsilon
\sum_{p=\pm}
v^2
\biggl \{
\gamma^+_{p L}(\varepsilon)
\,
\gamma^-_{p R}(\varepsilon)
\,
\cos^2 \frac{\phi}{2}
+
\gamma^+_{p L}(\varepsilon)
\,
\gamma^-_{\bar{p} R}(\varepsilon)
\,
\sin^2 \frac{\phi}{2}
\biggl \},
\label{eqn:ss}
\\
\Sigma^{LR}_{j \bar{j}}
&\simeq&
\frac{3 \pi \ri}{4}
\int \rd \varepsilon
\sum_{p=\pm}
\left[
\gamma^+_{p L}(\varepsilon)
\,
\gamma^-_{\bar{p} R}(\varepsilon-\Delta_{\bar{j} j})
\,
{\rm Re}
\left \{
1
+
\sigma_{1 \, p}(\varepsilon)
+
\sigma_{1 \, \bar{p}}(\varepsilon-\Delta_{\bar{j} j})
\right \}
\cos^2 \frac{\phi}{2}
\right.
\nonumber \\
&+&
\left.
\gamma^+_{p L}(\varepsilon)
\,
\gamma^-_{p R}(\varepsilon-\Delta_{\bar{j} j})
\,
{\rm Re}
\left \{
1
+
\sigma_{0 \, p}(\varepsilon)
+
\sigma_{0 \, \bar{p}}(\varepsilon-\Delta_{\bar{j} j})
\right \}
\sin^2 \frac{\phi}{2}
\right],
\label{eqn:st}
\end{eqnarray}
\end{widetext}
where we neglected the integral
$
{\rm Re} \int \rd \varepsilon' \gamma_p(\varepsilon')
/(\varepsilon+\ri \eta-\varepsilon') $. The subscript $j=1(0)$
denotes the total-spin quantum number and $\bar{j}=0(1)$. Here
$\Delta_{10}=-\Delta_{01}=J_{\rm RKKY}(\phi)$ and
$
\gamma_{p r}^{\pm}(\varepsilon)
=
\gamma_p (\varepsilon)
f^{\pm}(\varepsilon-\mu_r)
$
denotes the \lq\lq lesser" or \lq\lq greater" Green function
where $\mu_L=-\mu_R=eV/2$.
The function $\sigma_{0(1) p}$ defined by
\begin{equation}
\sigma_{0 (1) p}(\varepsilon)
=
\int \rd \varepsilon'
\frac{
\gamma^+_{L p}(\varepsilon')
+
\gamma^+_{R \bar{p} (p)}(\varepsilon')
}
{\varepsilon+\ri \eta-\varepsilon'},
\label{eqn:g}
\end{equation}
gives the logarithmic divergence related with Kondo correlations.
By substituting Eq. (\ref{eqn:green}) to Eq. (\ref{eqn:g}), we
obtain
\begin{eqnarray}
\sigma_{0 \pm}(\varepsilon)
\!\!
&\simeq&
\!\!
2 \bar{J}
\ln \frac{2 {\rm e}^\gamma D}{\pi T},
\nonumber
\\
\sigma_{1 \pm}(\varepsilon)
\!\!
&\simeq&
\!\!
\sigma_{0 \pm}(\varepsilon)
\pm
2 \bar{J}
\,
{\rm Re}
\biggl [
{\rm e}^{\ri k_{\rm F} l}
\biggl \{
\ln
\frac{2 T^*}{\pi T}
\nonumber \\
&+&
{\rm Ei} \left( -\ri k_{\rm F} l \right)
\biggl \}
\biggl ],
\nonumber
\end{eqnarray}
for $V=0$ and $\varepsilon \ll T, T^*$. Here ${{\rm Ei}(x)}$
denotes the exponential integral function and $\gamma \approx
0.577$ is the Euler constant.
Equation (\ref{eqn:defRKKY}) supplemented with Eq. (\ref{eqn:chi})
and Eqs. (\ref{eqn:tt}), (\ref{eqn:ss}) and (\ref{eqn:st}) are
main results of this paper.

Using the partial self-energy, Eqs. (\ref{eqn:tt}), (\ref{eqn:ss})
and (\ref{eqn:st}), the current can be expressed as
\begin{equation}
I
=
-
\frac{\ri e}{\hbar}
\sum_{j,j'=0,1}
P_j \,
\left \{
\Sigma^{L R}_{j \, j'}
-
(L \leftrightarrow R)
\right \},
\label{eqn:current}
\end{equation}
where probabilities $ P_0$ for a singlet state and $P_1$ for each
of particular triplet states (we consider no Zeeman splitting),
can be obtained for the linear response from the Boltzmann
distribution as
\begin{equation}
P_0=\frac{1}{1+3 \, \exp(-\beta J_{\rm RKKY}(\phi))}, \; \;
P_1=\frac{1-P_0}{3} \; .
 \label{eqn:probability}
\end{equation}
 The linear conductance is defined as $G = \lim_{V \rightarrow 0}
\partial I/ \partial V$.

\section{results and discussion}

For the AB ring geometry without quantum dots in arm, the
conductance oscillates as a function of the flux $\phi$
\cite{Gefen}. Furthermore, because of the orbital phase, the
conductance also oscillates as a function of the length of the arm
$l$ for enough low temperatures: As the thermal
excitation of lead electrons scrambles various orbital phases, the
oscillatory component would be reduced for temperatures above the
characteristic temperature $T^*$, where characteristic length $h
v_{\rm F}/T$ reaches $l$.
When one local spin, i.e. QD, is embedded in each arm of the
AB ring,
in addition to the oscillatory component,
the non-oscillatory background of oscillations
related to spin-flip processes will appear:
Spin-flip processes do not contribute to
the interference effect \cite{Akera} because if a local spin is
flipped, we can determine the path which an electron propagated.
Such non-oscillatory background
reduces the portion of oscillatory component.
However, if we take account of the RKKY interaction,
the oscillatory component
can be enhanced from the following mechanism:
First, according to Eq.~(\ref{eqn:probability}),
the probabilities for singlet $P_0$ and
triplet $P_1$ states will be affected by
the RKKY coupling constant $J_{\rm RKKY}(\phi)$,
which is a oscillatory function in terms of $\phi$ and $l$.
Second, as the conductance would be sensitive to a state
of local spins,
it would show also the oscillatory behavior
related to the oscillations of $J_{\rm RKKY}(\phi)$.
Such RKKY dominant oscillations one could expect for the enough
low temperature $T \! \ll \! |J_{\rm RKKY}(0)| \! \ll \! T^*$.

In the following, we will discuss the properties of
our system for temperatures
where the thermal scrambling of orbital phases
is unimportant $T \! \ll \! T^*$
and above the Kondo temperature $T \! \gg \! T_{\rm K}$.
We note that as $|J_{\rm RKKY}(\phi)|  \! \ll \! T^*$, the
modification of orbital phases by inelastic spin-flip scattering
events is also unimportant.

\subsection{ $l$ - dependence}

\begin{figure}[ht]
\includegraphics[width=0.7\columnwidth]{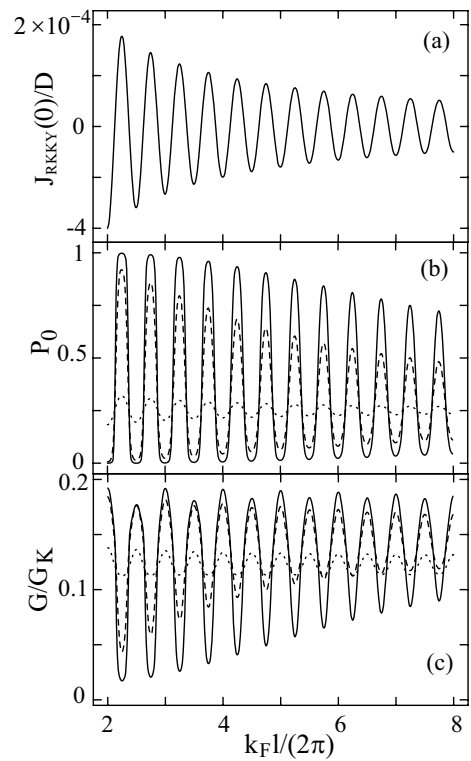}
\caption{Length dependent (a) RKKY coupling constant,
(b) probability for singlet state
and
(c) conductance
for
$T/D=5 \times 10^{-5}$ (solid line),
$10^{-4}$ (dashed line)
and
$10^{-3}$ (dotted line).
Parameters are taken as
$\phi=0$ and $\bar{J}=0.04$.
}
\label{fig:G2}
\end{figure}

First we will discuss the RKKY oscillations without magnetic flux
$\phi=0$ as a function of the distance $l$ between two QDs.
 Figure \ref{fig:G2}(a) shows the RKKY oscillations of the
coupling constant $J_{\rm RKKY}(0)$ as a function of the length of
an electron path $l$.
 It oscillates with the period of $ k_F l/\pi=1$, and shows local minima
at integer values of $k_{\rm F} l/\pi $ corresponding to
ferromagnetic (F) coupling and local maxima at half-integer values
of $k_{\rm F} l/\pi$ corresponding to the antiferromagnetic (AF)
coupling between the spins. The amplitude of the oscillations
decay with $~1/k_{\rm F} l$ as predicted for RKKY interaction in
quasi 1D geometry \cite{Yafet}.
In Figure \ref{fig:G2}(b), there is a plot of the probability
$P_0$ of the singlet state. At low temperature $T \lesssim |J_{\rm
RKKY}(0)|$ (the solid line), a singlet state (triplet state) is
formed when value of $k_{\rm F} l/\pi$ is close to half-integer
(integer).
 As the temperature increases (the dashed line for $T \sim |J_{\rm
RKKY}(0)|$ and the dotted line for $T \gg |J_{\rm RKKY}(0)|$) the
amplitude of oscillations is suppressed and system approaches
uniform distribution between the singlet and triplet states - $P_0
= P_0 = 1/4$.
There are also the oscillations of the conductance [Fig.
\ref{fig:G2}(c)] with the period of $ k_{\rm F} l/\pi = 1$. In the
same way as in Fig. \ref{fig:G2}(b), the amplitude of oscillations
is suppressed for $T \gg |J_{\rm RKKY}(0)|$, what indicates that
in regime $T < |J_{\rm RKKY}(0)|$ the conductance oscillations are
mainly determined by the RKKY interaction.
Experimentally, it can be difficult to control the length of arms
keeping other parameters fixed. However, the conductance
oscillations would be possible to observe by changing the Fermi
wave number $k_{\rm F}$, by controlling the carrier density of 2D
electron gas (2DEG) with an additional gate.

\subsection{ $\phi$ - dependence }

Though above discussions suggest that the RKKY interaction
dominates the length dependent conductance, it would be more
convenient experimentally to measure the flux dependence. In the
following we will discuss the modification of AB conductance
oscillations by the presence of RKKY interaction.

As we mentioned before by means of external flux $ \phi$ one can
change the amplitude of the RKKY interaction but not its sign
since $( 2 + 2 \cos \phi) \geq 0$. In the particular experimental
situation depending on the length $l$  of the arm and Fermi wave
vector $ k_{\rm F} $ the spins can be coupled ferromagnetically or
antiferromagnetically. By means of flux $ \phi$ one can control
the strength of the interaction but does not switch between them.
For this reason it is generic to discuss three typical situations,
for which we are able to get analytic results. These three cases
are classified by the value of the RKKY coupling constant: (i) the
uncorrelated local spins case ($\left|J_{\rm RKKY}(\phi)\right| \!
\ll \! T$), (ii) the ferromagnetic coupling case ($-J_{\rm
RKKY}(\phi) \! \gg \! T$) and (iii) the antiferromagnetic coupling
($J_{\rm RKKY}(\phi) \! \gg \! T$).

(i) Uncorrelated local spins limit is realize for high temperature
$\left|J_{\rm RKKY}(\phi)\right| \ll T$ or for the flux $\phi
\approx \pi + 2 \pi n$ since then, according to
Eq.~(\ref{eqn:defRKKY}), the RKKY interaction is weak, $ J_{\rm
RKKY}(\phi) \rightarrow 0 $.
 In this case (i), the
local-spin state is distributed with equal probability among a
singlet state and a triplet state $P_1 \approx P_0 \approx 1/4$
[see Eq.~(\ref{eqn:probability})]. The conductance is expressed as
\begin{eqnarray}
\frac{G}{G_{\rm K}}
&\simeq&
(\pi \bar{J})^2
\biggl \{
v^2
\left(
1
+
\cos \phi
\,
\cos^2 k_{\rm F} l
\right)
\nonumber \\
&+&
3
\left (
1+4 \bar{J}
\ln \frac{2 {\rm e}^{\gamma} D}{\pi T}
\right )
\biggl \},
\label{eqn:approx1}
\end{eqnarray}
where $G_{\rm K}=e^2/h$ is the quantum conductance.
The first term, which is proportional to $v^2$ and thus
independent of spin-flip processes, is attributed to the phase
coherent component of the cotunneling process. It shows the
ordinary AB oscillations.
The second term in Eq. (\ref{eqn:approx1}), which is related to
spin-flip processes (does not depend on $\phi$), form the
background of AB oscillations. We can see that with decreasing of
temperature the Kondo correlations enhance the background: The
second term can be interpreted as the parallel conductance through
two independent spin-1/2 local moments whose conductance is
enhanced by Kondo correlations \cite{note1}.
In the third order contribution in $\bar{J}$ in  Eq.
(\ref{eqn:approx1}), there is no interference related to the
orbital phase $k_{\rm F} l$, which was pointed out by Beal-Monod
\cite{BealMonod}. We explicitly showed by Eq. (\ref{eqn:approx1})
that there is also no interference related to the AB phase in the
third order contribution.

(ii) The ferromagnetic coupling for $-J_{\rm RKKY}(\phi) \gg T$ :
In this case, two local spins form a triplet state $P_1 \approx
1/3$ and $P_0 \approx 0$
 [see
Eq. (\ref{eqn:probability})]. Thus, the conductance is that of
$S=1$ Kondo model plus the potential scattering. For the case of
long distance between QDs   ($k_{\rm F} l \gg 1$),
\begin{eqnarray}
\frac{G}{G_{\rm K}}
&\simeq&
2
(\pi \bar{J})^2
\left[
4 \bar{J}
\cos^2 k_{\rm F} l \,
\cos^2 \frac{\phi}{2} \,
\ln \frac{2 T^*}{\pi T}
\right.
\nonumber \\
&+&
\left.
\left(
1+\frac{v^2}{2}+
2 \bar{J}
\ln \frac{2 {\rm e}^{\gamma} D}{\pi T}
\right)
\left(
1+\cos \phi \, \cos^2 k_{\rm F} l
\right)
\right].
\nonumber \\
\label{eqn:approx2}
\end{eqnarray}
For the opposite case, $k_{\rm F} l \ll 1$,
we obtain the same equation as Eq. (\ref{eqn:approx2})
with replacing $T^*$ in the
logarithm by ${\rm e}^{-\gamma} D$.
The striking feature is that as opposed to the case (i), the Kondo
correlations enhance the oscillatory component as it is shown in
the second term of Eq. (\ref{eqn:approx2}). Loosely speaking, two
spins are no longer independent phase-breaking scatterers because
they \lq \lq observe" each other and the Kondo correlations
enhances the AF coupling of each QD spin to the conducting
electrons spins.
The first term of Eq. (\ref{eqn:approx2}) shows the logarithmic
divergence, whose the cutoff energy is equal to the characteristic
temperature of the orbital phase coherence $T^*$. This term
appears because the spin-1 moment stretches over $l$.
Using Eq. (\ref{eqn:approx2}), we can relate the F coupling of
spins by RKKY interaction [Fig. \ref{fig:G2}(a)] with the maximum
in the conductance [Fig. \ref{fig:G2}(c)] around integer values of
$k_{\rm F} l/\pi$.

(iii) The antiferromagnetic coupling, $J_{\rm RKKY}(\phi) \! \gg
\! T$: In this case two local moments form a singlet state $P_1 \!
\approx \! 0$ and $P_0 \! \approx \! 1$ [see Eq.
(\ref{eqn:probability})]. As the singlet state is decoupled form
lead electrons, i.e. electrons flowing through QDs cannot excite
local spins to a triplet state, so only the potential scattering
process contributes to the conductance:
\begin{equation}
G/G_{\rm K}
\simeq
(\pi \bar{J})^2
\,
v^2
(
1+\cos \phi \, \cos^2 k_{\rm F} l
).
\label{eqn:approx3}
\end{equation}
Because we consider the Coulomb blockade regime, the cotunneling
current is very small. It is the reason why the conductance is
suppressed around each half integer value of $k_{\rm F} l/\pi$
[Fig. \ref{fig:G2}(c)], where the RKKY coupling is
antiferromagnetic [Fig. \ref{fig:G2}(a)].
Here we note that situation (ii) and (iii) are not realized for
the flux $\phi \approx \pi + 2 \pi n$ since  $J_{\rm RKKY}(\phi)$
is small there and the limit (i) is approached. For $ | J_{\rm
RKKY}(0) | > T $ by means of the flux $ \phi $ we can tune between
(ii) and (i), or (iii) and (i) but not between (ii) and (iii)
situations.

\begin{figure}[ht]
\includegraphics[width=1.0\columnwidth]{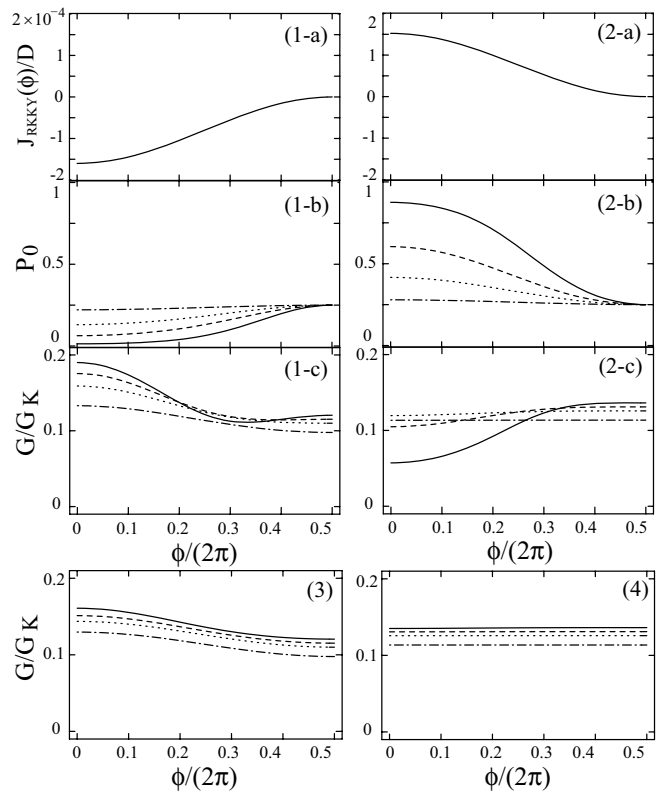}
\caption{Flux dependent RKKY coupling constant ((1-a) and (2-a)),
probability for singlet state((1-b) and (2-b))
and
conductance ((1-c) and (2-c))
for
$\bar{J}=0.04$.
Panels (1-a), (1-b) and (1-c)
correspond to the F coupling case
($k_{\rm F} l/(2 \pi) = 5$)
and
panels (2-a), (2-b) and (2-c)
correspond to AF coupling case
($k_{\rm F} l/(2 \pi) = 5.25$).
Flux dependent conductance for
(3) $k_{\rm F} l/(2 \pi) = 50$ and
(4) $50.25$.
The solid, dashed, dotted, and dot-dashed lines
show the results for
$T/D=
5 \times 10^{-5},
10^{-4},
2 \times 10^{-4}$, and $10^{-3}$, respectively.
}
\label{fig:G1}
\end{figure}

For above three cases, we obtained simple analytic results and
clarify that the local-spin state due to RKKY interaction causes
the pronounced effect on the conductance.
Next, we will analyze the conductance of the system for the full
range of the flux $ \phi $ and discuss the additional structures
caused by the flux dependent RKKY interaction, which can be an
evidence of the RKKY interaction in our system.
 Figures
\ref{fig:G1}(1-a) and (2-a) show the RKKY coupling constant
$J_{\rm RKKY}(\phi)$ as a function of the flux $\phi/(2 \pi)$. The
former shows plot for F coupling case ($k_{\rm F} l/\pi$ is an
integer) and the latter show the plot for AF coupling case
($k_{\rm F} l/\pi$ is a half-integer). The panels (1-b) and (2-b)
are the corresponding plots of the probability for the singlet
state for various temperatures and the panels (1-c) and (2-c) are
plots of the conductance.
In the vicinity of zero flux $ \pi = 0 $, electron wave functions
constructively interfere and thus the maximum RKKY interaction is
induced [panels (1-a) and (2-a)]. For F coupling case, a triplet
state is formed, i.e. $P_0 \sim 0$, at low temperature [panel
(1-b)] and thus the conductance is enhanced [panel (1-c)] as
discussed in (ii). For AF coupling case at low temperature, a
singlet is formed [panel (2-b)] and the conductance is suppressed
[panel (2-c)] as it was discussed in (iii).
At half flux, electron wave functions destructively interfere and
the RKKY interaction is switched off [panels (1-a) and (2-a)].
Surprisingly at half flux we can observed the maximum in the
conductance for both situations F and AF. According to discussion
in (i), this maximum is caused by the term in
Eq.~(\ref{eqn:approx1}), which does not depend on the flux, and
which corresponds to incoherent transport thought the two
independent spin-1/2 local moments related to Kondo correlations.
Especially for AF coupling case, it leads to the effective phase
shift of AB conductance oscillations by $\pi$ [panels (2-c)].

In order to compare our results with the limit, where the RKKY
interaction is negligible, we show curves of AB oscillations for
$|J_{\rm RKKY}(\phi)| \ll T$ in panels (3) and (4). As discussed
in (i), the component of the ordinary AB oscillations is very
small. The Kondo correlations only enhance the background and they
do not promote characteristic structures as the case of the AF or
F coupling.

Here we note some features to distinguish experimentally the RKKY
dominant oscillations from the ordinary AB oscillations.
The first feature is the characteristic temperature below which
the oscillations can be observed: The characteristic temperature
of the ordinary AB oscillations $T^*$ is higher than that of RKKY
dominant oscillations $|J_{\rm RKKY}(0)|$ by the factor $\sim \!
\bar{J}^{-2}$. One can point out that the RKKY dominant
oscillations is sensitive to the temperature.
The second feature is the temperature dependence of the amplitude
of oscillations:
Suppose we decrease the temperature from enough
high temperature $T \gg |J_{\rm RKKY}(0)|$, where
singlet and triplet probabilities are $P_1 \! \approx P_1 \! \approx 1/4$
and the conductance
is expressed by Eq. (\ref{eqn:approx1}).
As temperature is lowered,
singlet and triplet probabilities are modified as
$P_0 \approx 1/4 \, (1-3 J_{\rm RKKY}(\phi)/(4T))$
and
$P_1 \approx 1/4 \, (1+J_{\rm RKKY}(\phi)/(4 T))$.
Therefore, the correction depending on both the orbital phase and
the AB phase
\begin{eqnarray}
\delta G
\!
&\simeq& 
\! \!
-
(\pi \bar{J})^2 
\biggl \{
3
\!
\left (
1+4 \bar{J} \ln \frac{2
{\rm e}^{\gamma} D}{\pi T} 
\right ) 
(2+\cos^2 k_{\rm F} l \cos \phi)
\nonumber \\
&+&
2 v^2
(1+\cos^2 k_{\rm F} l \cos \phi)
\biggl \}
\frac{J_{\rm RKKY}(\phi)}{4 T},
\nonumber
\end{eqnarray}
emerges. It grows as $\propto (\ln T)/T$; the logarithmic
correction is related to the Kondo correlations. We expect that
with the help of the Kondo correlations one can distinguish the
RKKY dominant AB conductance oscillations from the ordinary AB
oscillations.


Here we will note on the Onsager symmetry. For the two-terminal
geometry, it means that the conductance is an even function of the
flux. We can see that the RKKY coupling constant $J_{\rm
RKKY}(\phi)$ is an even function of the flux (Eq.
(\ref{eqn:defRKKY})). This property depends only on the symmetry
of the Hamiltonian under the inversion of time and magnetic field
\cite{Sta-Phy} and does not depends on the assumption of the
mirror symmetry.

%
%
Finally we discuss on the range of parameters.
For a 2 DEG system at an AlAs/GaAs heterostructure,
the carrier density of which is typically
$\approx 3.8 \times 10^{15} {\rm m}^{-2}$\cite{Kobayashi1},
the Fermi energy and the Fermi wave length are
$\varepsilon_{\rm F}
\!
\approx
\!
D
\!
\approx
\!
14 \, {\rm meV}$
and
$2 \pi/k_{\rm F} \! \approx \! 40 \, {\rm nm}$,
respectively.
The RKKY coupling constant $J_{\rm RKKY}(0) \!\! \sim \!\!
\bar{J}^2 D/(k_{\rm F} l)$ should be larger than the Kondo
temperature $T_{\rm K} \! \sim \! D \exp(-1/(2 \bar{J}))$,
$|J_{\rm RKKY}(0)| \! \gg \! T_{\rm K}$, otherwise, each spin-1/2
local moment forms Kondo singlet and are screened out and thus the
RKKY interaction is unimportant. In our calculations, we put
$\bar{J} =0.04$ which gives the small Kondo temperature, $T_{\rm
K} \sim 3.7 \times 10^{-6} D \ll |J_{\rm RKKY}(0)|$. Because we
adopted the perturbation theory, the temperature should be above
the Kondo temperature, $T \gg T_{\rm K}$. In order to obtain the
large RKKY interaction $J_{\rm RKKY}(0) \gtrsim T$ we put the size
of the ring as $k_{\rm F} l \approx 5 \times 2 \pi$ (about 200
nm).

For the case of a small AB ring, the Zeeman splitting could become
important. In order to reduce the Zeeman splitting $E_Z$ keeping
the number of fluxes constant, one could increase the size of the
AB ring because $J_{\rm RKKY}(0)/E_{\rm Z} \sim l$. In such a
case, the consideration of the Kondo regime $T \lesssim T_{\rm
K}$, could be needed because RKKY interaction was also reduced so
the limit $T_K \gtrsim |J_{\rm RKKY}(0)|$ was approached.


\section{summary}

In conclusion, we have theoretically investigated the RKKY
interaction acting between local spins, i.e. two QDs with odd
numbers of electrons in CB regime, embedded in the AB ring. We
assumed the parity symmetry of the system and such an assumption
does not change the result qualitatively. We calculated the RKKY
coupling constant and the conductance above the Kondo temperature,
$T \gg T_{\rm K}$. The RKKY coupling constant, the sign of which
oscillates as a function of the distance, also depends on the flux
and the distance between two QDs.
When the RKKY interaction is ferromagnetic,
two local spins form a triplet state around zero flux,
where the electron wave constructively interfere,
and thus the maximum RKKY interaction is induced.
As the temperature decreases, the amplitude of
AB oscillations is enhanced by Kondo correlations,
which is the distinctive difference between
the ordinary AB oscillations
and those of the ferromagnetically coupled two local spins.
The maximum was found at half flux
where the RKKY interaction is switched off and
the conductance is described by the parallel conductance
of two independent spin-1/2 local moments
whose conductance is enhanced by Kondo correlations.
When the RKKY interaction is AF,
the phase of AB oscillations is shifted by $\pi$.
It is because around zero flux, where we obtain
the maximum AF interaction,
two local spins form a singlet state,
which is decoupled from the lead electrons.

\begin{acknowledgements}

We would like to thank
V. Dugaev, H. Ebisawa, J. K{\"o}nig, S. Maekawa, T. P. Pareek and
G. Sch{\"o}n for valuable discussions and comments. H.I was
supported by MEXT, Grant-in-Aid for Scientific Research on the
Priority Area "Semiconductor Nanospintronics" No. 14076204.
\end{acknowledgements}

\appendix

\section{third order perturbation theory}
\label{appendix:perturbation}

\begin{figure}[ht]
\includegraphics[width=0.90\columnwidth]{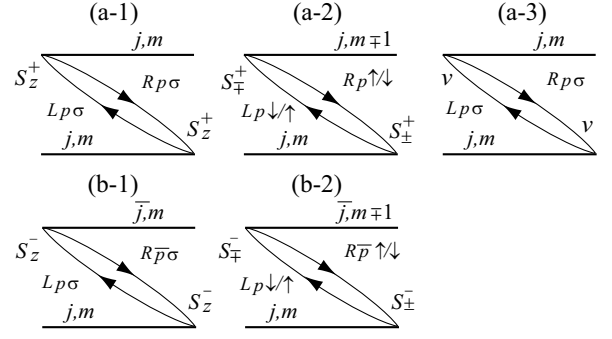}
\caption{ The diagrams for the second order partial self-energy
representing the transition preserving total-spin quantum number
$j$, [(a-1), (a-2) and (a-3)] and the singlet-triplet transition
[(b-1) and (b-2)]. Directed lines represent propagators for lead
electrons. Thick solid lines on the Keldysh contour (two
horizontal lines) represent propagators for the local spins. }
\label{fig:diagram1}
\end{figure}

In this Appendix, we present detailed calculations of the
third-order partial self-energy in terms of $\bar{J}$ on the basis
of the diagrammatic technique in the real-time domain
\cite{Schoeller_Schon,Konig1}.
Figure \ref{fig:diagram1} shows the second order diagrams for the
partial self-energy representing the transition preserving the
total-spin quantum number $j$, $\Sigma^{LR}_{jj}$ ((a-1), (a-2)
and (a-3)) and the singlet-triplet transition, $\Sigma^{LR}_{j
\bar{j}}$ ($j=0,1$) ((b-1) and (b-2)). Green functions of lead
electrons are represented by directed solid lines, which are also
called \lq\lq reservoir lines", and solid lines on the Keldysh
contour (two horizontal lines) represents propagators of local
spins.
Here diagrams (a-1) and (a-3)
represent different processes.
For the former case,
we must count factor $-1$ for the vertex denoted with
$S^{+}_z$ when $\sigma=\downarrow$.
We omitted diagrams which could be obtained by applying the mirror
rule \cite{Konigdiplom}.

Following the rules in Ref.~\cite{Schoeller_Schon}, the diagram
(a-1) plus its mirror diagram can be calculated as
\begin{eqnarray}
{\Sigma^{LR}_{jj}}^{\text{(a-1)}}
\!\!\!
&=&
\!\!\!
\sum_{\stackrel{\scriptstyle{p=\pm}}{-j \leq m \leq j}}
\!\!\!\!
\frac{\ri \pi}{8}
\int \rd \varepsilon
\gamma^+_{p L}(\varepsilon)
\gamma^-_{p R}(\varepsilon)
\cos^2 \! \frac{\phi}{2}
\nonumber \\
&\times&
{\rm Re} \,
\langle j,m|
2 {S^+_z}^2
|j,m \rangle.
\label{eqn:ttzero}
\end{eqnarray}
The results for the diagrams (a-2) and (a-3), which we term
${\Sigma^{LR}_{jj}}^{\text{(a-2)}}$ and
${\Sigma^{LR}_{jj}}^{\text{(a-3)}}$, can be obtained from
Eq.~(\ref{eqn:ttzero}) by changing $2 {S^+_z}^2$ to $S^+_{\pm}
S^+_{\mp}$ and to $2 v^2$, respectively. In the same way, the
diagram (b-1) plus its mirror diagram is calculated as
\begin{eqnarray}
{\Sigma^{LR}_{j \bar{j}}}^{\text{(b-1)}}
\!\!\!
&=&
\!\!\!
\sum_{\stackrel{\scriptstyle{p=\pm}}{-j \leq m \leq j}}
\!\!\!
\frac{\ri \pi}{8}
\int \rd \varepsilon
\gamma^+_{p L}(\varepsilon)
\gamma^-_{\bar{p} R}(\varepsilon-\Delta_{j \bar{j}})
\cos^2 \!\! \frac{\phi}{2}
\nonumber \\
&\times&
{\rm Re} \,
\langle j,m|
2 {S^-_z}^2
|j,m \rangle,
\label{eqn:stzero}
\end{eqnarray}
where $\Delta_{j \bar{j}}$ is the energy difference between the
total-spin quantum number $j$ state and $\bar{j}$ state. The
result for the diagram (b-2), which we denote by ${\Sigma^{LR}_{j
\bar{j}}}^{\text{(b-2)}}$, is obtained from Eq.~(\ref{eqn:stzero})
by replacing of $2 {S^-_z}^2$ by $S^+_{\pm} S^+_{\mp}$.

\begin{figure}[ht]
\includegraphics[width=0.90\columnwidth]{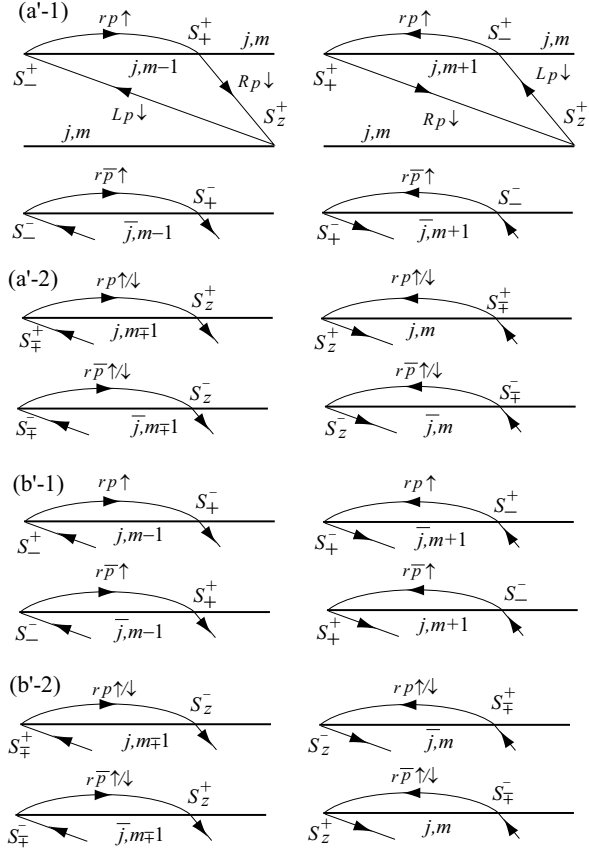}
\caption{
The third order diagrams:
Each four diagrams of (a'-1), (a'-2), (b'-1) and (b'-2)
show corrections for the vertex on the upper branch of
the diagram (a-1), (a-2), (b-1) and (b-2)
in Fig. \ref{fig:diagram1},
respectively.
}
\label{fig:diagram2}
\end{figure}

The third order diagrams give the vertex correction to the second
order diagrams. Figures \ref{fig:diagram2}(a'-1), (a'-2), (b'-1)
and (b'-2) show the correction for the vertex on the upper branch
of diagrams (a-1), (a-2), (b-1) and (b-2) in
Fig.~\ref{fig:diagram1}, respectively. Except for the topmost two
diagrams, we omitted the lower branch of each diagram, which is
exactly the same as for the corresponding diagram in Fig.
\ref{fig:diagram1}.
The left diagrams and the right diagrams show direct tunneling
processes and exchange processes, respectively.
We did not show the correction for the diagram (a-3) because it
is proportional to
$\sum_{-j \leq m \leq j} \langle j,m| S^+_z |j,m \rangle$
and thus vanishes.
For example, the topmost two diagrams in Fig. \ref{fig:diagram2}
plus their mirror diagrams are calculated
by utilizing
the commutation relations Eq. (\ref{eqn:commute})
as
\begin{eqnarray}
& & \!\!\! {\Sigma^{LR}_{jj}}^{\text{(a-1) correction}}=
\nonumber \\
& &
\frac{2 \ri}{4^3} \,
{\rm Im}
\sum_{r \, p}
\int
\rd \varepsilon_1
\rd \varepsilon_2
\rd \varepsilon_3
\frac{
\gamma^+_{L p}(\varepsilon_1)
\gamma^-_{R p}(\varepsilon_3)
}{\varepsilon_1-\varepsilon_3+\ri \eta}
\cos^2 \frac{\phi}{2}
\nonumber \\
&\times&
\!\!\!
\langle j,m|
\,
S^+_z
\left \{
\frac{\gamma^-_{r p}(\varepsilon_2) \, S_+^+  S_-^+}
{\varepsilon_1-\varepsilon_2+\ri \eta}
-
\frac{\gamma^+_{r p}(\varepsilon_2) \, S_-^+  S_+^+}
{\varepsilon_2-\varepsilon_3+\ri \eta} \right \} |j,m \rangle
\nonumber \\ &\simeq& \frac{\ri \pi}{8} \sum_{p} \int \rd
\varepsilon \gamma^+_{L p}(\varepsilon) \gamma^-_{R
p}(\varepsilon) \cos^2 \frac{\phi}{2} \nonumber \\ &\times& \!\!\!
{\rm Re} \left \{ \frac{\sigma_{1 p}(\varepsilon)}{4} \langle j,m|
2{S^+_z}^2 |j,m \rangle \right \}, \label{eqn:example}
\end{eqnarray}
where $\sigma_{1 p}$ is defined in Eq. (\ref{eqn:g}).
Here we counted
the minus sign for a loop with three vertices
in the anticlockwise direction
and
we dropped terms except for the renormalization
of the transmission probability.
We checked that terms which we dropped are canceled out
by the other diagrams than those depicted in Fig. \ref{fig:diagram2},
i.e. diagrams in which the position of a lower vertex
is inbetween upper two vertices.
Further we neglected the integral
$
{\rm Re}
\int \rd \varepsilon'
\gamma_p(\varepsilon')
/(\varepsilon+\ri \eta-\varepsilon')
$
which is at most
$\sim \!\!
\bar{J} \varepsilon/D$
for
$k_{\rm_F} l \!\! \ll \!\! 1$
or
$\sim \!\!
\bar{J}
\{ |\varepsilon|/D+1/(k_{\rm_F} l) \}$
for
$k_{\rm_F} l \!\! \gg \!\! 1$.
By adding Eq. (\ref{eqn:ttzero}) and Eq. (\ref{eqn:example}),
we obtain Eq. (\ref{eqn:ttzero})
with replacing $2 {S^+_z}^2$ by $2 (1+\sigma_{1p}(\varepsilon)/4) {S^+_z}^2$.

Other two diagrams in panel (a'-1) can be calculated in the same
way.
In Fig. \ref{fig:diagram2} we omitted diagrams obtained by
reversing direction and indices for spin and lead of reservoir
lines. By calculation of all such diagrams and adding them to
Eq.~(\ref{eqn:ttzero}), we obtain Eq.~(\ref{eqn:ttzero}) with
replacing $2 {S^+_z}^2$ by $2 z_{jj}' {S^+_z}^2$, where the
renormalization factor is given by
\begin{equation}
z_{jj}'
=
1+
\frac{1}{2}
\left \{
\sigma_{1 p}(\varepsilon)
+
\frac{
\sigma_{1 \bar{p}}(\varepsilon-\Delta_{\bar{j} j})
+
\sigma_{1 \bar{p}}(\varepsilon-\Delta_{j \bar{j}})
}{2}
\right \}.
\nonumber
\end{equation}
In Fig. \ref{fig:diagram2} we did not show the
lower vertex corrections, which are given in the
same way as the upper vertex corrections.
By counting lower vertex corrections, $z_{jj}'$ is modified as
\begin{equation}
z_{jj}
=
1+
\sigma_{1 p}(\varepsilon)
+
\frac{
\sigma_{1 \bar{p}}(\varepsilon-\Delta_{\bar{j} j})
+
\sigma_{1 \bar{p}}(\varepsilon-\Delta_{j \bar{j}})
}{2}.
\label{eqn:a1}
\end{equation}
Finally,
the third order contributions change
${\Sigma^{LR}_{jj}}^{\text{(a-1)}}$
to
${\Sigma^{LR}_{jj}}^{\text{(a-1)}(2)}$,
where the latter
is obtained from the former by
replacing $2 {S^+_z}^2$ with $2 z_{jj} {S^+_z}^2$.
For the other diagrams than those of panel (a'-1), we can repeat
the same discussions as above. The result for diagrams (a-2) and
(a'-2) and their derivative diagrams, which we term
${\Sigma^{LR}_{jj}}^{\text{(a-2)}(2)}$, is obtained from Eq.
(\ref{eqn:ttzero}) by replacing $2 {S_z^+}^2$ with $z_{jj}
S^+_{\pm} S^+_{\mp}$.

By calculating diagrams in (b'-1) and their derivative diagrams,
and adding them to the diagram (b-1), we obtain ${\Sigma^{LR}_{j
\bar{j}}}^{\text{(b-1)}(2)}$, which is the same expression as Eq.
(\ref{eqn:stzero}) with replacing $2 {S^-_z}^2$ by $2 z_{j
\bar{j}}{S^-_z}^2$. Here
\begin{equation}
z_{j \bar{j}}
=
1+
\sigma_{1 p}(\varepsilon)
+
\sigma_{1 \bar{p}}(\varepsilon-\Delta_{\bar{j} j}).
\label{eqn:b1}
\end{equation}
%
The result for the diagrams (b-2) and (b'-2)
and their derivative diagrams,
which we term
${\Sigma^{LR}_{j \bar{j}}}^{\text{(b-2)}(2)}$
is obtained from Eq. (\ref{eqn:stzero})
by replacing
$2 {S_z^-}^2$ with $z_{j \bar{j}} S^-_{\pm} S^-_{\mp}$.
Finally, by summarizing ${\Sigma^{LR}_{jj}}^{\text{(a-1)}(2)}$,
${\Sigma^{LR}_{jj}}^{\text{(a-2)}(2)}$ and
${\Sigma^{LR}_{jj}}^{\text{(a-3)}}$, we obtain the first term of
Eq. (\ref{eqn:tt}) for $j\!=\!1$ and the first term of Eq.
(\ref{eqn:ss}) for $j\!=\!0$. By adding ${\Sigma^{LR}_{j
\bar{j}}}^{\text{(b-1)}(2)}$ to ${\Sigma^{LR}_{j
\bar{j}}}^{\text{(b-2)}(2)}$, we obtain the first term of Eq.
(\ref{eqn:st}).

For now, we have explained only the diagrams related to the
time-reversal symmetric term, corresponding to the first and
second lines in Eq. (\ref{eqn:symmetrizedHT}). Diagrams related to
the time-reversal symmetry breaking term, the third line in Eq.
(\ref{eqn:symmetrizedHT}), are obtained from Figs.
\ref{fig:diagram1} and \ref{fig:diagram2} by changing the parity
indices of the right reservoir lines.
For example, the corresponding diagram of (a-1)
is calculated as
\begin{eqnarray}
{\Sigma^{LR}_{jj}}^{\widetilde{\text{(a-1)}}}
\!\!\!
&=&
\!\!\!
\sum_{\stackrel{\scriptstyle{p=\pm}}{m=\pm1,0}}
\!\!\!\!
\frac{\ri \pi}{8}
\int \rd \varepsilon
\gamma^+_{p L}(\varepsilon)
\gamma^-_{\bar{p} R}(\varepsilon)
\sin^2 \! \frac{\phi}{2}
\nonumber \\
&\times&
{\rm Re} \,
\langle j,m|
2 {S^+_z}^2
|j,m \rangle.
\nonumber
\end{eqnarray}
%
For vertex corrections, the change in the parity indices of the
right reservoir lines corresponds to the operation of the
replacement of $\sigma_{1 \, p}$ by $\sigma_{0 \, p}$.
Thus the second terms
of Eqs. (\ref{eqn:tt}), (\ref{eqn:ss}) and (\ref{eqn:st})
can be obtained from first terms by replacing
$\sigma_{1 \, p}$,
$\cos^2 (\phi/2)$
and
$\gamma^-_{p(\bar{p}) R}$
with
$\sigma_{0 \, p}$,
$\sin^2 (\phi/2)$
and
$\gamma^-_{\bar{p}(p) R}$, respectively.


\begin{thebibliography}{29}
\expandafter\ifx\csname natexlab\endcsname\relax\def\natexlab#1{#1}\fi
\expandafter\ifx\csname bibnamefont\endcsname\relax
  \def\bibnamefont#1{#1}\fi
\expandafter\ifx\csname bibfnamefont\endcsname\relax
  \def\bibfnamefont#1{#1}\fi
\expandafter\ifx\csname citenamefont\endcsname\relax
  \def\citenamefont#1{#1}\fi
\expandafter\ifx\csname url\endcsname\relax
  \def\url#1{\texttt{#1}}\fi
\expandafter\ifx\csname urlprefix\endcsname\relax\def\urlprefix{URL }\fi
\providecommand{\bibinfo}[2]{#2}
\providecommand{\eprint}[2][]{\url{#2}}

\bibitem[{\citenamefont{Kittel}(1968)}]{Kittel}
\bibinfo{author}{\bibfnamefont{C.}~\bibnamefont{Kittel}}
  (\bibinfo{publisher}{Academic Press}, \bibinfo{address}{New York and London},
  \bibinfo{year}{1968}), Solid State Physics, p.~\bibinfo{pages}{1}.

\bibitem[{\citenamefont{Maekawa and Shinjo}(2002)}]{maekawaBook}
\bibinfo{editor}{\bibfnamefont{S.}~\bibnamefont{Maekawa}} \bibnamefont{and}
  \bibinfo{editor}{\bibfnamefont{T.}~\bibnamefont{Shinjo}}, eds.,
  \emph{\bibinfo{title}{Spin Dependent Transport in Magnetic Nanostructures}},
  ADVANCES IN CONDENSED MATTER SCIENCE (\bibinfo{publisher}{Taylor \& Francis},
  \bibinfo{address}{London and New York}, \bibinfo{year}{2002}).

\bibitem[{\citenamefont{Bruno}(1995)}]{Bruno}
\bibinfo{author}{\bibfnamefont{P.}~\bibnamefont{Bruno}},
  \bibinfo{journal}{Phys. Rev. B} \textbf{\bibinfo{volume}{52}},
  \bibinfo{pages}{411} (\bibinfo{year}{1995}).

\bibitem[{\citenamefont{Holleitner et~al.}(2001)\citenamefont{Holleitner,
  Decker, Qin, Eberl, and Blick}}]{Holleitner}
\bibinfo{author}{\bibfnamefont{A.~W.} \bibnamefont{Holleitner}},
  \bibinfo{author}{\bibfnamefont{C.~R.} \bibnamefont{Decker}},
  \bibinfo{author}{\bibfnamefont{H.}~\bibnamefont{Qin}},
  \bibinfo{author}{\bibfnamefont{K.}~\bibnamefont{Eberl}}, \bibnamefont{and}
  \bibinfo{author}{\bibfnamefont{R.~H.} \bibnamefont{Blick}},
  \bibinfo{journal}{Phys. Rev. Lett.} \textbf{\bibinfo{volume}{87}},
  \bibinfo{pages}{256802} (\bibinfo{year}{2001}).

\bibitem[{\citenamefont{Yacoby et~al.}(1995)\citenamefont{Yacoby, Heiblum,
  Mahalu, and Shtrikman}}]{Yacoby}
\bibinfo{author}{\bibfnamefont{A.}~\bibnamefont{Yacoby}},
  \bibinfo{author}{\bibfnamefont{M.}~\bibnamefont{Heiblum}},
  \bibinfo{author}{\bibfnamefont{D.}~\bibnamefont{Mahalu}}, \bibnamefont{and}
  \bibinfo{author}{\bibfnamefont{H.}~\bibnamefont{Shtrikman}},
  \bibinfo{journal}{Phys. Rev. Lett.} \textbf{\bibinfo{volume}{74}},
  \bibinfo{pages}{4047} (\bibinfo{year}{1995}).

\bibitem[{\citenamefont{Kobayashi et~al.}(2002)\citenamefont{Kobayashi, Aikawa,
  Katsumoto, and Iye}}]{Kobayashi1}
\bibinfo{author}{\bibfnamefont{K.}~\bibnamefont{Kobayashi}},
  \bibinfo{author}{\bibfnamefont{H.}~\bibnamefont{Aikawa}},
  \bibinfo{author}{\bibfnamefont{S.}~\bibnamefont{Katsumoto}},
  \bibnamefont{and} \bibinfo{author}{\bibfnamefont{Y.}~\bibnamefont{Iye}},
  \bibinfo{journal}{Phys. Rev. Lett.} \textbf{\bibinfo{volume}{88}},
  \bibinfo{pages}{256806} (\bibinfo{year}{2002}).

\bibitem[{\citenamefont{Loss and Sukhorukov}(2000)}]{Loss}
\bibinfo{author}{\bibfnamefont{D.}~\bibnamefont{Loss}} \bibnamefont{and}
  \bibinfo{author}{\bibfnamefont{E.~V.} \bibnamefont{Sukhorukov}},
  \bibinfo{journal}{Phys. Rev. Lett.} \textbf{\bibinfo{volume}{84}},
  \bibinfo{pages}{1035} (\bibinfo{year}{2000}).

\bibitem[{\citenamefont{Georges and Meir}(1999)}]{Georges}
\bibinfo{author}{\bibfnamefont{A.}~\bibnamefont{Georges}} \bibnamefont{and}
  \bibinfo{author}{\bibfnamefont{Y.}~\bibnamefont{Meir}},
  \bibinfo{journal}{Phys. Rev. Lett.} \textbf{\bibinfo{volume}{82}},
  \bibinfo{pages}{3508} (\bibinfo{year}{1999}).

\bibitem[{\citenamefont{K{\"o}nig and Gefen}(2001)}]{Konig_1}
\bibinfo{author}{\bibfnamefont{J.}~\bibnamefont{K{\"o}nig}} \bibnamefont{and}
  \bibinfo{author}{\bibfnamefont{Y.}~\bibnamefont{Gefen}},
  \bibinfo{journal}{Phys. Rev. Lett.} \textbf{\bibinfo{volume}{86}},
  \bibinfo{pages}{3855} (\bibinfo{year}{2001}).

\bibitem[{\citenamefont{K{\"o}nig and Gefen}(2002)}]{Konig_2}
\bibinfo{author}{\bibfnamefont{J.}~\bibnamefont{K{\"o}nig}} \bibnamefont{and}
  \bibinfo{author}{\bibfnamefont{Y.}~\bibnamefont{Gefen}},
  \bibinfo{journal}{Phys. Rev. B} \textbf{\bibinfo{volume}{65}},
  \bibinfo{pages}{045316} (\bibinfo{year}{2002}).

\bibitem[{\citenamefont{Bruder et~al.}(1996)\citenamefont{Bruder, Fazio, and
  Schoeller}}]{Bruder}
\bibinfo{author}{\bibfnamefont{C.}~\bibnamefont{Bruder}},
  \bibinfo{author}{\bibfnamefont{R.}~\bibnamefont{Fazio}}, \bibnamefont{and}
  \bibinfo{author}{\bibfnamefont{H.}~\bibnamefont{Schoeller}},
  \bibinfo{journal}{Phys. Rev. Lett.} \textbf{\bibinfo{volume}{76}},
  \bibinfo{pages}{114} (\bibinfo{year}{1996}).

\bibitem[{\citenamefont{Gefen et~al.}(1984)\citenamefont{Gefen, Imry, and
  Azbel}}]{Gefen}
\bibinfo{author}{\bibfnamefont{Y.}~\bibnamefont{Gefen}},
  \bibinfo{author}{\bibfnamefont{Y.}~\bibnamefont{Imry}}, \bibnamefont{and}
  \bibinfo{author}{\bibfnamefont{M.~Y.} \bibnamefont{Azbel}},
  \bibinfo{journal}{Phys. Rev. Lett.} \textbf{\bibinfo{volume}{52}},
  \bibinfo{pages}{129} (\bibinfo{year}{1984}).

\bibitem[{\citenamefont{Jayaprakash et~al.}(1981)\citenamefont{Jayaprakash,
  Krishna-murthy, and Wilkins}}]{Jayaprakash}
\bibinfo{author}{\bibfnamefont{C.}~\bibnamefont{Jayaprakash}},
  \bibinfo{author}{\bibfnamefont{H.~R.} \bibnamefont{Krishna-murthy}},
  \bibnamefont{and} \bibinfo{author}{\bibfnamefont{J.~W.}
  \bibnamefont{Wilkins}}, \bibinfo{journal}{Phys. Rev. Lett.}
  \textbf{\bibinfo{volume}{47}}, \bibinfo{pages}{737} (\bibinfo{year}{1981}).

\bibitem[{\citenamefont{Hackenbroich and
  Weidenm{\"u}ller}(1996)}]{Hackenbroich}
\bibinfo{author}{\bibfnamefont{G.}~\bibnamefont{Hackenbroich}}
  \bibnamefont{and} \bibinfo{author}{\bibfnamefont{H.~A.}
  \bibnamefont{Weidenm{\"u}ller}}, \bibinfo{journal}{Phys. Rev. Lett.}
  \textbf{\bibinfo{volume}{76}}, \bibinfo{pages}{110} (\bibinfo{year}{1996}).

\bibitem[{\citenamefont{Beal-Monod}(1969)}]{BealMonod}
\bibinfo{author}{\bibfnamefont{M.~T.} \bibnamefont{Beal-Monod}},
  \bibinfo{journal}{Phys. Rev.} \textbf{\bibinfo{volume}{178}},
  \bibinfo{pages}{874} (\bibinfo{year}{1969}).

\bibitem[{\citenamefont{Jones and Varma}(1987)}]{Jones1}
\bibinfo{author}{\bibfnamefont{B.~A.} \bibnamefont{Jones}} \bibnamefont{and}
  \bibinfo{author}{\bibfnamefont{C.~M.} \bibnamefont{Varma}},
  \bibinfo{journal}{Phys. Rev. Lett.} \textbf{\bibinfo{volume}{58}},
  \bibinfo{pages}{843} (\bibinfo{year}{1987}).

\bibitem[{\citenamefont{Jones et~al.}(1988)\citenamefont{Jones, Varma, and
  Wilkins}}]{Jones2}
\bibinfo{author}{\bibfnamefont{B.~A.} \bibnamefont{Jones}},
  \bibinfo{author}{\bibfnamefont{C.~M.} \bibnamefont{Varma}}, \bibnamefont{and}
  \bibinfo{author}{\bibfnamefont{J.~W.} \bibnamefont{Wilkins}},
  \bibinfo{journal}{Phys. Rev. Lett.} \textbf{\bibinfo{volume}{61}},
  \bibinfo{pages}{125} (\bibinfo{year}{1988}).

\bibitem[{\citenamefont{Jones et~al.}(1989)\citenamefont{Jones, Kotliar, and
  Millis}}]{Jones3}
\bibinfo{author}{\bibfnamefont{B.~A.} \bibnamefont{Jones}},
  \bibinfo{author}{\bibfnamefont{B.~G.} \bibnamefont{Kotliar}},
  \bibnamefont{and} \bibinfo{author}{\bibfnamefont{A.~J.}
  \bibnamefont{Millis}}, \bibinfo{journal}{Phys. Rev. B}
  \textbf{\bibinfo{volume}{39}}, \bibinfo{pages}{3415} (\bibinfo{year}{1989}).

\bibitem[{\citenamefont{Schwabe et~al.}(1996)\citenamefont{Schwabe, Elliott,
  and Wingreen}}]{Schwabe}
\bibinfo{author}{\bibfnamefont{N.~F.} \bibnamefont{Schwabe}},
  \bibinfo{author}{\bibfnamefont{R.~J.} \bibnamefont{Elliott}},
  \bibnamefont{and} \bibinfo{author}{\bibfnamefont{N.~S.}
  \bibnamefont{Wingreen}}, \bibinfo{journal}{Phys. Rev. B}
  \textbf{\bibinfo{volume}{54}}, \bibinfo{pages}{12953} (\bibinfo{year}{1996}).

\bibitem[{\citenamefont{Kubala and K{\"o}nig}(2003)}]{kubala_2}
\bibinfo{author}{\bibfnamefont{B.}~\bibnamefont{Kubala}} \bibnamefont{and}
  \bibinfo{author}{\bibfnamefont{J.}~\bibnamefont{K{\"o}nig}},
  \bibinfo{journal}{Phys. Rev. B} \textbf{\bibinfo{volume}{67}},
  \bibinfo{pages}{205303} (\bibinfo{year}{2003}).

\bibitem[{\citenamefont{Yafet}(1987)}]{Yafet}
\bibinfo{author}{\bibfnamefont{Y.}~\bibnamefont{Yafet}},
  \bibinfo{journal}{Phys. Rev. B} \textbf{\bibinfo{volume}{36}},
  \bibinfo{pages}{3948} (\bibinfo{year}{1987}).

\bibitem[{\citenamefont{Matho and Beal-Monod}(1972)}]{Matho}
\bibinfo{author}{\bibfnamefont{K.}~\bibnamefont{Matho}} \bibnamefont{and}
  \bibinfo{author}{\bibfnamefont{M.~T.} \bibnamefont{Beal-Monod}},
  \bibinfo{journal}{Phys. Rev. B} \textbf{\bibinfo{volume}{5}},
  \bibinfo{pages}{1899} (\bibinfo{year}{1972}).

\bibitem[{\citenamefont{Schoeller and Sch{\"o}n}(1994)}]{Schoeller_Schon}
\bibinfo{author}{\bibfnamefont{H.}~\bibnamefont{Schoeller}} \bibnamefont{and}
  \bibinfo{author}{\bibfnamefont{G.}~\bibnamefont{Sch{\"o}n}},
  \bibinfo{journal}{Phys. Rev. B} \textbf{\bibinfo{volume}{50}},
  \bibinfo{pages}{18436} (\bibinfo{year}{1994}).

\bibitem[{\citenamefont{K{\"o}nig et~al.}(1996)\citenamefont{K{\"o}nig, Schmid,
  Schoeller, and Sch{\"o}n}}]{Konig1}
\bibinfo{author}{\bibfnamefont{J.}~\bibnamefont{K{\"o}nig}},
  \bibinfo{author}{\bibfnamefont{J.}~\bibnamefont{Schmid}},
  \bibinfo{author}{\bibfnamefont{H.}~\bibnamefont{Schoeller}},
  \bibnamefont{and}
  \bibinfo{author}{\bibfnamefont{G.}~\bibnamefont{Sch{\"o}n}},
  \bibinfo{journal}{Phys. Rev. B} \textbf{\bibinfo{volume}{54}},
  \bibinfo{pages}{16820} (\bibinfo{year}{1996}).

\bibitem[{\citenamefont{Akera}(1993)}]{Akera}
\bibinfo{author}{\bibfnamefont{H.}~\bibnamefont{Akera}},
  \bibinfo{journal}{Phys. Rev. B} \textbf{\bibinfo{volume}{47}},
  \bibinfo{pages}{6835} (\bibinfo{year}{1993}).

\bibitem[{not()}]{note1}
\bibinfo{note}{Equation (\ref{eqn:approx1}) recovers a previous result for
  single impurity \cite{Kaminski}. Namely, Eq. (\ref{eqn:approx1}) with
  dropping the factor $\ln (2 {\rm e}^\gamma/\pi)$ and also the flux and length
  dependent term is twice as large as Eq. (27) in Ref. \cite{Kaminski}, which
  corresponds to the transport throught two independent impurities. In our
  notations $\bar{J}$ and $D$ correspond to $\nu {\cal J}^{(0)}_{rr'}$
  ($r,r'\!\! = \!\! L,R$) and $D_0$ in Ref. \cite{Kaminski}, respectively.}

\bibitem[{\citenamefont{Kubo et~al.}(1985)\citenamefont{Kubo, Toda, and
  Hashitsume}}]{Sta-Phy}
\bibinfo{author}{\bibfnamefont{R.}~\bibnamefont{Kubo}},
  \bibinfo{author}{\bibfnamefont{M.}~\bibnamefont{Toda}}, \bibnamefont{and}
  \bibinfo{author}{\bibfnamefont{N.}~\bibnamefont{Hashitsume}},
  \emph{\bibinfo{title}{Statistical Physics II}}, vol.~\bibinfo{volume}{31} of
  \emph{\bibinfo{series}{Springer Series in Solid-State Sciences}}
  (\bibinfo{publisher}{Springer-Verlag}, \bibinfo{address}{Berlin Heidelberg
  New York Tokyo}, \bibinfo{year}{1985}).

\bibitem[{\citenamefont{K{\"o}nig}(unpublished)}]{Konigdiplom}
\bibinfo{author}{\bibfnamefont{J.}~\bibnamefont{K{\"o}nig}},
  \bibinfo{journal}{diploma thesis (University of Karlsruhe, 1995)}
  (\bibinfo{year}{unpublished}).

\bibitem[{\citenamefont{Kaminski et~al.}(2000)\citenamefont{Kaminski, Nazarov,
  and Glazman}}]{Kaminski}
\bibinfo{author}{\bibfnamefont{A.}~\bibnamefont{Kaminski}},
  \bibinfo{author}{\bibfnamefont{{\rm Yu}.~V.} \bibnamefont{Nazarov}},
  \bibnamefont{and} \bibinfo{author}{\bibfnamefont{L.~I.}
  \bibnamefont{Glazman}}, \bibinfo{journal}{Phys. Rev. B}
  \textbf{\bibinfo{volume}{62}}, \bibinfo{pages}{8154} (\bibinfo{year}{2000}).

\end{thebibliography}

\end{document}